\newif\ifpdf
\ifx\pdfoutput\undefined
   \pdffalse
\else
   \pdfoutput=1
   \pdftrue
\fi

\ifpdf
\documentclass[12pt,a4paper]{article}
\else
\documentclass[12pt,a4paper, dvips]{article}
\fi

\usepackage{latexsym}
\usepackage{amssymb}
\usepackage{amsmath}
\usepackage[authoryear]{natbib}

\ifpdf
    \usepackage[pdftex]{graphicx}
    \usepackage{color}
     \definecolor{myred}{rgb}{0.5,0,0}
     \definecolor{myblue}{rgb}{0,0,0.75}
     \definecolor{mygreen}{rgb}{0,0.5,0}
   \usepackage[pdftex,
               pdftitle={Calculating credit risk capital charges with the one-factor model},
               pdfsubject={Credit portfolio risk},
               pdfkeywords={One-factor model; capital charge; granularity adjustment;
quantile derivative},
               pdfauthor={Susanne Emmer, Dirk Tasche},
               pdfstartview=FitBH,
               breaklinks=true,
               colorlinks=true,
               citecolor=myred,
               linkcolor=myblue,
               urlcolor=mygreen]{hyperref}
\else
    \usepackage[dvips]{graphicx}
   \usepackage{hyperref}
\fi

\newtheorem{theorem}{Theorem}[section]
\newtheorem{lemma}[theorem]{Lemma}
\newtheorem{remark}[theorem]{Remark}

\newtheorem{proposition}[theorem]{Proposition}
\newtheorem{definition}[theorem]{Definition}
\newtheorem{corollary}[theorem]{Corollary}
\newtheorem{assumption}[theorem]{Assumption}

\numberwithin{equation}{section}

\begin{document}

\renewcommand{\thepage}{\roman{page}}

\title{Calculating credit risk capital charges with the one-factor model%
}

\author{Susanne Emmer\thanks{Dr.~Nagler \& Company GmbH,
                   Maximilianstra\ss{}e 47, 80538 M\"unchen, Germany\newline E-mail:
                   susanne.emmer@nagler-company.com}\hspace{3cm}
Dirk Tasche\thanks{Deutsche Bundesbank, Postfach 10 06 02, 60006 Frankfurt am
Main, Germany\newline E-mail: tasche@ma.tum.de}\ %
\thanks{The opinions expressed in this note are those of the
authors and do not necessarily reflect views of the Deutsche Bundesbank
or of Nagler \& Company.} }

\date{}
\maketitle

\newpage

\textbf{Summary}\\ Even in the simple one-factor credit portfolio model that
underlies the Basel~II regulatory capital rules coming into force in 2007, the exact
contributions to credit value-at-risk can only be calculated with Monte-Carlo
simulation or with approximation algorithms that often involve numerical integration. 
As this may require a lot of computational time, there is a need
for approximate analytical formulae. In this note, we develop formulae
according to two different approaches: the granularity adjustment approach
initiated by
M.~Gordy and T.~Wilde, and a
semi-asymptotic approach. The application of the formulae is illustrated with
a numerical example.\\[1ex]

\newpage

\setcounter{page}{1}
\renewcommand{\thepage}{\arabic{page}}
\section{Introduction}

In the past two years, one-factor models\footnote{See \citet{FMc2002} for an
overview of credit modeling methodologies.} for credit portfolio risk have
become popular. On the one side, they can rather readily be handled from a
computational point of view, and, in particular, allow to avoid lengthy Monte
Carlo simulations which in general are needed when dealing with more
sophisticated credit risk models. On the other side, they are capable to
incorporate simple dependence structures which are to a certain degree
sufficient for a rudimentary form of credit risk management.

Often, one-factor models admit a decomposition of the portfolio loss variable
into a monotonic function of the factor and a residual. The former part of
the decomposition is called \emph{systematic} risk whereas the latter part
is called \emph{specific} or \emph{idiosyncratic} risk. Sometimes, the
portfolio loss variable converges in some sense to a monotonic function
of the factor. This observation can be used as point of departure for
analytic approximations of important statistics of the portfolio like
quantiles of the portfolio loss variable \emph{(value-at-risk, VaR)}.

\citet{Gordy01} was the first to suggest this approach which he called
\emph{granularity adjustment} for the one-factor
model credit portfolio that
underlies the Basel~II regulatory capital rules coming into force in 2007 
\citep[see][]{BC04}. 
Then it was refined by \citet{Wilde01}\footnote{%
The authors thank Tom Wilde for stimulating the writing of this paper and for
useful comments on its first version.} and \citet{Pykhtin02}. Only
recently, \citet{MartinWilde02} observed that the results by \citet{GLS00}
make feasible an easier and more systematic way to derive the adjustments. In
the following (Section~\ref{sec:gran}), we  reexamine the granularity
adjustment in the one-factor model and develop new formulae 
for computing adjustments at the transaction level (Corollary~\ref{co:1}). Additionally, in
Section~\ref{sec:part} we present a new alternative \emph{semi-asymptotic} approach to capital
charges at transaction level which relies on a limiting procedure applied to a
part of the portfolio only. In contrast to the granularity adjustment
approach, the semi-asymptotic approach is able to capture fully the effects of exposure
concentrations on the risk contributions of single transactions.
Section~\ref{sec:num} provides a numerical
example. We conclude with a short summary in Section
\ref{sec:concl}.

\section{The granularity adjustment approach}
\label{sec:gran}

In order to explain the approach we will follow, we consider two random
variables $L$ and $X$. $L$ denotes the portfolio loss whereas $X$ reflects the
value of an economic factor which causes the dependence between the different
transactions building up the portfolio. The conditional expectation of $L$
given $X$, $\mathrm{E}[L\,|\,X]$, is considered the systematic part of the
portfolio loss. This can be motivated by the decomposition
\begin{equation}
  \label{eq:1}
  L \,=\, \mathrm{E}[L\,|\,X] + L - \mathrm{E}[L\,|\,X]
\end{equation}
which leads to
\begin{align}
  \label{eq:2}
\begin{split}
  \mathrm{var}[L]&= \mathrm{E}\bigl[\mathrm{var}[L
\,|\,X]\bigr] + \mathrm{var}\bigl[\mathrm{E}[L\,|\,X]\bigr]\\
&= \mathrm{E}\bigl[\mathrm{var}[L - \mathrm{E}[L\,|\,X]\,\big|\,X]\bigr] +
\mathrm{var}\bigl[\mathrm{E}[L\,|\,X]\bigr],
\end{split}
\end{align}
where $\mathrm{var}$ denotes variance and $\mathrm{var}[\,\cdot\,|\,X]$ means
conditional variance. Under the assumption that conditional on the value of
the economic factor $X$ the elementary transactions in the portfolio are
independent, the term $\mathrm{E}\bigl[\mathrm{var}[L \,|\,X]\bigr]$ in
(\ref{eq:2}) is likely to converge to zero with a growing number of
transactions in the portfolio. Therefore, the specific risk $ L -
\mathrm{E}[L\,|\,X]$ of the portfolio may be considered \emph{incremental} and
small compared to the systematic risk $\mathrm{E}[L\,|\,X]$. Note that by the
factorization lemma $\mathrm{E}[L\,|\,X]$ may be written as $g(X)$ with
some appropriately chosen function $g$. This becomes interesting when $g$
turns out to be monotonic since in this case values for some statistics of
$\mathrm{E}[L\,|\,X]$ like quantiles can be easily computed from the
corresponding statistics of $X$. We fix this as a formal assumption.
\begin{assumption}\label{as:1}
The conditional expectation of $L$ given the factor $X$ can be written as
\begin{equation*}
  \mathrm{E}[L\,|\,X]\,=\,g(X),
\end{equation*}
where $g$ is continuous and strictly increasing or decreasing.
The distribution of $X$ is continuous, i.e.\ $\mathrm{P}[X = x] = 0$ for all $x$.
\end{assumption}
For $\alpha\in (0,1)$ and any random variable $Y$, define the $\alpha$-quantile of $Y$ by
\begin{equation}
  \label{eq:4}
  q_\alpha(Y)\,=\,\inf\bigl\{ y\in\mathbb{R}: \mathrm{P}[Y\le y]\ge\alpha\bigr\}.
\end{equation}
As $L$ is interpreted as a portfolio loss variable, we are interested in high
quantiles of $L$, i.e.\ in the case of $\alpha$ being close to $100\%$ like
$\alpha = 99.9\%$ as in Basel~II. Then the value of the quantile can be
considered the amount of capital necessary to keep the portfolio holder
solvent with a very high probability. In this context, $q_\alpha(L)$ is also
called \emph{value-at-risk} at level $\alpha$ of the portfolio loss.

The granularity adjustment approach to the calculation of the
$\alpha$-quantile $q_\alpha(L)$ of the portfolio loss is essentially a second
order Taylor expansion. Observing that $L = \mathrm{E}[L\,|\,X] + h\,(L
  -\mathrm{E}[L\,|\,X])$ for $h=1$, this can be seen with the following short calculation:
\begin{align}
  q_\alpha(L) &= q_\alpha\bigl(\mathrm{E}[L\,|\,X] + h\,(L
  -\mathrm{E}[L\,|\,X])\bigr)\Big|_{h=1}\notag\\
&\approx q_\alpha\bigl(\mathrm{E}[L\,|\,X]\bigr) + \frac{\partial q_\alpha}{\partial
  h} \bigl(\mathrm{E}[L\,|\,X] + h\,(L
  -\mathrm{E}[L\,|\,X])\bigr)\Big|_{h=0}\label{eq:5} \\
& \qquad +\ \frac12 \,\frac{\partial^2 q_\alpha}{\partial
  h^2} \bigl(\mathrm{E}[L\,|\,X] + h\,(L
  -\mathrm{E}[L\,|\,X])\bigr)\Big|_{h=0}.\notag
\end{align}
The derivatives of $q_\alpha\bigl(\mathrm{E}[L\,|\,X] + h\,(L
  -\mathrm{E}[L\,|\,X])\bigr)$ with respect to $h$ can be calculated thanks to
results by \citet{GLS00} \citep[see][for the derivatives of arbitrary large
orders]{MartinWilde02b}.
This way, we obtain
\begin{align}
  \frac{\partial q_\alpha}{\partial
  h} \bigl(\mathrm{E}[L\,|\,X] + h\,(L
  -\mathrm{E}[L\,|\,X])\bigr)\Big|_{h=0} &=
  \mathrm{E}\bigl[L-\mathrm{E}[L\,|\,X]\,\big|\,\mathrm{E}[L\,|\,X] =
  q_\alpha(\mathrm{E}[L\,|\,X])\bigr]\notag\\
& = 0,\label{eq:6}
\end{align}
where (\ref{eq:6}) follows from Assumption \ref{as:1}, since $
q_\alpha(\mathrm{E}[L\,|\,X]) = g\bigl(q_\alpha(X)\bigr)$ if $g$ is
increasing, and  $q_\alpha(\mathrm{E}[L\,|\,X]) =
g\bigl(q_{1-\alpha}(X)\bigr)$ if $g$ is decreasing. Under Assumption
\ref{as:1}, $\mathrm{E}[L\,|\,X]$ has a density if $X$ has a density $h_X$. If
 this density of $\mathrm{E}[L\,|\,X]$ is denoted by $\gamma_L$, it can be
 determined according to
\begin{align}\label{eq:dens}
\gamma_L(y)&=h_X(g^{-1}(y))\, \left|g'(g^{-1}(y))\right|^{-1}.
\end{align}
With regard to the second derivative in (\ref{eq:5}), we obtain by means of
the corresponding result of \citet{GLS00}
\begin{align}\label{eq:7}
  \frac{\partial^2 q_\alpha}{\partial
  h^2} \bigl(\mathrm{E}[L\,|\,X] + h\,(L
  -\mathrm{E}[L\,|\,X])\bigr)\Big|_{h=0} &= - \frac{\partial
  \mathrm{var}}{\partial x}\bigl(L\,|\,X =g^{-1}(x)\bigr)\Big|_{x=g(q_{\beta}(X))} \\[1ex]
& \qquad -\ \frac{\mathrm{var}\bigl[L\,|\,X
  =q_{\beta}(X)\bigr]}{\gamma_L\bigl(g(q_{\beta}(X))\bigr)}\,
\frac{\partial \gamma_L}{\partial x}(x)\Big|_{x=g(q_{\beta}(X))},\notag
\end{align}
with $\beta = \alpha$, if $g$ is increasing, and $\beta = 1- \alpha$ if $g$ is decreasing.

In order to be able to carry out numerical calculations by means of
(\ref{eq:5}) and (\ref{eq:7}), we have to fix a stochastic model, the
one-factor model in our case. Define the indicator variable $\mathbf{1}_D$ of
the event $D$ by 
\begin{equation}
  \label{eq:ind}
\mathbf{1}_D(\omega) \ =\ 
\begin{cases}
  1, & \omega \in D,\\
  0, & \omega \notin D.
\end{cases}
\end{equation}
We then define the portfolio loss $L_n$ by
\begin{subequations}
 \begin{align}
  \label{eq:8}
  L_n & =  L_n(u_1, \ldots, u_n)\,=\, \sum_{i=1}^n u_i\,
\mathbf{1}_{D_i},\\
D_i & = \{\sqrt{\rho_i}\,X+\sqrt{1-\rho_i}\,\xi_i\le c_i\},\label{eq:8a}
\end{align}
\end{subequations}
where $u_i\ge 0$, $i = 1, \ldots, n$,
denotes the weight or the exposure of asset $i$ in the portfolio, $0 < \rho_i
< 1$ and $c_i \ge 0$, $i = 1, \ldots, n$, are constants, and $X, \xi_1,
\ldots, \xi_n$ are independent random variables with continuous distributions.
As mentioned above, $X$ is interpreted as an economic factor that influences
all the assets in the portfolio but to different extents. The so-called
\emph{asset correlation} $\rho_i$ measures the degree of the $i$-th asset's
exposure to the systematic risk expressed by $X$. The random variables $\xi_i$
are assumed to model the idiosyncratic (or specific) risk of the assets. The
event $D_i$ given by \eqref{eq:8a} can then be interpreted as the event that
default occurs with the $i$-th asset because its value falls below some critical threshold.

It is easy to show that in this case Assumption \ref{as:1} holds with
decreasing $g = g_n$ where $g_n(x)$ is given by
\begin{equation}
  \label{eq:9}
  g_n(x)\,=\,\sum_{i=1}^n u_i\,\mathrm{P}\bigl[\xi_i\le
\textstyle{\frac{c_i-\sqrt{\rho_i}\,x}{\sqrt{1-\rho_i}}}].
\end{equation}
(\ref{eq:8}) implies that
\begin{equation}
  \label{eq:9a}
  \mathrm{var}[L_n\,|\,X=x]\,=\,\sum_{i=1}^n u_i^2\,\mathrm{P}
\bigl[\xi_i\le
\textstyle{\frac{c_i-\sqrt{\rho_i}\,x}{\sqrt{1-\rho_i}}}\bigr]\,
\bigl(1-\mathrm{P}
\bigl[\xi_i\le
\textstyle{\frac{c_i-\sqrt{\rho_i}\,x}{\sqrt{1-\rho_i}}}\bigr]\bigr).
\end{equation}
Under the additional assumption that $\inf_i c_i > -\infty$ and $\sup_i \rho_i
< 1$, it follows from (\ref{eq:9a}) that in case of i.i.d.\  $\xi_1, \xi_2,
\ldots$ and $X$ we have \addtocounter{footnote}{1}
\begin{subequations}
\begin{align}\label{eq:var_lim}
\lim_{n\to\infty} \mathrm{E}\bigl[\mathrm{var}[L_n\,|\,X]\bigr] &= 0\\
\intertext{if and only
if\footnotemark[\thefootnote]}
\lim_{n\to\infty} \sum_{i=1}^n u_i^2 & =
0.\label{eq:cond}
\end{align}
\end{subequations}
\footnotetext[\thefootnote]{%
Of course, here we admit an additional dependence of $u_i$ on $n$, i.e.\ $u_i%
= u_{i,n}$.}%
We will see below (Remark \ref{rm:10}) that small values of $\sum_{i=1}^n
u_i^2$ correspond to the case where the portfolio is well diversified in the
sense that there are no essential exposure concentrations.

By (\ref{eq:var_lim}) and (\ref{eq:cond}), in case of large portfolios the
systematic part $\mathrm{var}\bigl[\mathrm{E}[L_n\,|\,X]\bigr]$ of the loss
variance $\mathrm{var}[L_n]$ (cf.\ (\ref{eq:2})) is the essential part of the
variance. Hence, in this case approximating $L_n$ by $\mathrm{E}[L_n\,|\,X]$
seems reasonable. This observation suggests approximating $q_\alpha(L_n)$ with
the right-hand side of (\ref{eq:5}), where $\mathrm{E}[L_n\,|\,X]$ is given by
$g_n(X)$ with $g_n(x)$ defined by (\ref{eq:9}).

Let us now specify the distributions of $X, \xi_1, \xi_2, \ldots, \xi_n$ as
all being the standard normal distributions. This implies
\begin{subequations}
  \begin{align}
    g_n(x) &= \sum_{i=1}^n u_i\,\Phi\bigl( \frac{c_i -
    \sqrt{\rho_i}\,x}
{\sqrt{1-\rho_i}} \bigr)\label{eq:10}\\
\intertext{and}
\mathrm{var}\bigl[L_n\,|\,X=x\bigr] &= \sum_{i=1}^n u_i^2\,\left(\Phi\bigl( \frac{c_i -
    \sqrt{\rho_i}\,x}
{\sqrt{1-\rho_i}} \bigr) - \Phi\bigl( \frac{c_i -
    \sqrt{\rho_i}\,x}
{\sqrt{1-\rho_i}} \bigr)^2\right)\label{eq:11}
  \end{align}
\end{subequations}
where $\Phi$ denotes the distribution function of the standard normal
distribution.

Since $g_n$ is decreasing, we obtain from (\ref{eq:dens}) and (\ref{eq:10})
for the density $\gamma_{L_n}$ of $\mathrm{E}[L_n\,|\,X]$
\begin{subequations}
  \begin{align}
 \gamma_{L_n}(x) &= -
 \frac{\phi\bigl(g_n^{-1}(x)\bigr)}{g'_n\bigl(g_n^{-1}(x)\bigr)} \label{eq:12}\\
\intertext{and}
 \frac{\partial \gamma_{L_n}}{\partial x}(x) &=
\frac{\phi\bigl(g_n^{-1}(x)\bigr)\,g''_n\bigl(g_n^{-1}(x)\bigr) -
\phi'\bigl(g_n^{-1}(x)\bigr)\,g'_n\bigl(g_n^{-1}(x)\bigr)}
{g'_n\bigl(g_n^{-1}(x)\bigr)^3}, \label{eq:13}
  \end{align}
\end{subequations}
with $\phi(x) = (\sqrt{2\,\pi})^{-1} e^{- 1/2\,x^2}$ denoting the density of
the standard normal distribution.

We are now in a position to specialize (\ref{eq:5}) for the case of normally
distributed underlying random variables. Plugging in the expressions from
(\ref{eq:11}), (\ref{eq:12}), and (\ref{eq:13}) into (\ref{eq:7}) and then the
resulting expression for the second derivative of the quantile into
(\ref{eq:5}) yields the following formula.
\begin{proposition}\label{pr:0}
Let $L_n$ be the portfolio loss variable as defined in (\ref{eq:8}). Assume
that the economic factor $X$ and the idiosyncratic risk factors $\xi_1,
\ldots, \xi_n$ are independent and standard normally distributed. Fix a
confidence level $\alpha \in (0,1)$. Then the second order Taylor
approximation of the quantile $q_\alpha(L_n)$ in the sense of (\ref{eq:5}) can
be calculated according to
\begin{eqnarray}
q_{\alpha}(L_n) & \approx &
g_n\bigl(q_{1-\alpha}(X)\bigr) \notag  \\
& + & \bigg( \sum_{i=1}^n u_i^2\,{\textstyle\sqrt{\frac{\rho_i}{1-\rho_i}}}\,
\phi\bigl( {\textstyle\frac{c_i -
    \sqrt{\rho_i}\,q_{1-\alpha}(X)}
{\sqrt{1-\rho_i}}} \bigr)
\bigl(1 - 2\,\Phi\bigl( {\textstyle\frac{c_i -
    \sqrt{\rho_i}\,q_{1-\alpha}(X)}
{\sqrt{1-\rho_i}}}
 \bigr)\bigr)\notag\\
& & 
+\,\Big(q_{1-\alpha}(X) + {\textstyle\frac{g''_n\bigl(q_{1-\alpha}(X)\bigr)}
{g'_n\bigl(q_{1-\alpha}(X)\bigr)}}\Big) \sum_{i=1}^n u_i^2\, \bigl(\Phi\bigl(
{\textstyle\frac{c_i -
    \sqrt{\rho_i}\,q_{1-\alpha}(X)}
{\sqrt{1-\rho_i}}} \bigr) - \Phi\bigl({\textstyle \frac{c_i -
    \sqrt{\rho_i}\,q_{1-\alpha}(X)}
{\sqrt{1-\rho_i}}} \bigr)^2\bigr)\bigg)\notag \\
& & \quad \times \,\bigl(2\,g'_n(q_{1-\alpha}(X))\bigr)^{-1}, \label{eq:14}
\end{eqnarray}
with $\phi$ and $\Phi$ denoting the density and the distribution function
respectively of the standard normal distribution.
\end{proposition}
\begin{remark}[Herfindahl index]\label{rm:10}
Consider the following special case of (\ref{eq:14}):
\begin{equation*}
  \rho_i = \rho, \quad\text{and}\quad c_i = c,\quad i = 1, \ldots, n, \quad
  \sum_{i=1}^n u_i = 1.
\end{equation*}
Here the $u_i$ have to be interpreted as the relative weights of the assets in
the portfolio. (\ref{eq:14}) then reads
\begin{eqnarray}\label{eq:equal}
q_{\alpha}(L_n) & \approx &
{\textstyle\Phi\left(\frac{c-\sqrt{\rho}\,q_{1-\alpha}(X)}{\sqrt{1-\rho}}\right)
- \left( 2\,\sqrt{\frac{\rho}{1-\rho}}\,
\phi\bigl(\frac{c-\sqrt{\rho}\,q_{1-\alpha}(X)}{\sqrt{1-\rho}}\bigr)\right)^{-1}}
\sum_{i=1}^n u_i^2
\notag\\
&  &\ \times \ {\textstyle \bigg(\sqrt{\frac{\rho}{1-\rho}}\,
\phi\bigl(\frac{c-\sqrt{\rho}\,q_{1-\alpha}(X)}{\sqrt{1-\rho}}\bigr)
\bigl(1-2\,\Phi\bigl(\frac{c-\sqrt{\rho}\,q_{1-\alpha}(X)}{\sqrt{1-\rho}}\bigr)\bigr)}\\
& & {\textstyle \quad + \ \bigl(q_{1-\alpha}(X) -
\sqrt{\frac{\rho}{1-\rho}}\,\frac{c-\sqrt{\rho}\,q_{1-\alpha}(X)}{\sqrt{1-\rho}}\bigr)
\Phi\left(\frac{c-\sqrt{\rho}\,q_{1-\alpha}(X)}{\sqrt{1-\rho}}\right)
\Phi\left(\frac{\sqrt{\rho}\,q_{1-\alpha}(X)-c}{\sqrt{1-\rho}}\right) \bigg).
}\notag
\end{eqnarray}
The case $u_i = 1/n$ for all $i$ is particularly important as it describes a
completely homogeneous portfolio without concentrations. If the $u_i$ are not
all equal, $\bigl(\sum_{i=1}^n u_i^2\bigr)^{-1}$ is approximately the
equivalent portfolio size that would render the portfolio completely
homogeneous. Assume that the $i$-th asset has absolute exposure $v_i$. Then
its relative weight $u_i$ is given by
\begin{equation*}
  u_i \,=\, \frac{v_i}{\sum_{j=1}^n v_j}.
\end{equation*}
In this case the above introduced equivalent portfolio size reads
$\dfrac{\bigl(\sum_{j=1}^n v_j\bigr)^2}{\sum_{j=1}^n v_j^2}$. This quantity is
known as \emph{Herfindahl index}. \citet{Gordy01} suggested the Herfindahl
index as a key constituent for constructing equivalent homogeneous portfolios.
(\ref{eq:equal}) yields strong support for this suggestion.
\end{remark}
Recall that in Proposition \ref{pr:0} not only $q_{\alpha}(L_n) =
q_\alpha\bigl(L_n(u_1,\ldots,u_n)\bigr)$ is a function of the weights
$u_1,\ldots, u_n$ but that also the function $g_n$ depends on $u_1,\ldots,
u_n$ via (\ref{eq:10}). A closer inspection of (\ref{eq:14}) reveals that the
right-hand side of the equation (as function of the weight vector $(u_1,
\ldots, u_n)$) is positively homogeneous of order 1. By Euler's theorem on the
representation of homogeneous functions as a weighted sum of the partial
derivatives, we can obtain canonical approximate capital charges on
transaction or sub-portfolio level which add up to the approximate
value-at-risk \citep[cf.][]{L96, Tasche99}. The capital charges will thus be
approximations to the terms $u_j\,\frac{\partial q_\alpha(L_n)}{\partial
u_j}$, $j = 1, \ldots, n$.

These approximate capital charges do not enjoy the portfolio invariance
which is the great advantage of the asymptotic capital charges suggested by
\citet{Gordy01}. Portfolio invariance in Gordy's sense means that the capital
charges of the single assets depend upon the characteristics of the assets
under consideration only but not upon the portolio composition. In particular,
when dealing with portfolio invariant capital charges, it does not really matter whether
an asset takes on $50\%$ of the total portfolio exposure or $0.05\%$ only,
because the capital charge would increase only linearly. However, in order to
capture exposure concentration effects, the capital charge of an asset
should grow more than linearly with the exposure as soon as the relative
weight of the asset in the portfolio exceeds some critical value.
Therefore, as portfolio invariance has the effect to entail
capital charges which are not sensitive to concentrations in the portfolio,
the partial derivatives approach has got its own attractiveness.
\begin{corollary}\label{co:1}
The partial derivative with respect to $u_j$ of the approximate quantile\\
$q_{\alpha}(L_n)=q_\alpha\bigl(L_n(u_1,\ldots,u_n)\bigr)$ as provided by the
right-hand side of (\ref{eq:14}) is given by
\begin{eqnarray}
\dfrac{\partial q_\alpha(L_n)}{\partial u_j}& \approx &
\Phi\left(\dfrac{c_j-\sqrt{\rho_j}\,q_{1-\alpha}(X)}{\sqrt{1-\rho_j}}\right)\notag\\
& +& \bigl(2\,g_n'(q_{1-\alpha}(X))^2\bigr)^{-1} \bigg(
2\,u_j\,A_j\,g_n'(q_{1-\alpha}(X)) -B_j \sum_{i=1}^nu_i^2 A_i \notag\\
& & +\ \bigl( E_j - \dfrac{B_j\,g_n''(q_{1-\alpha}(X))}{g_n'(q_{1-\alpha}(X))}
\bigr) \sum\limits_{i=1}^{n}u_i^2\,D_i\\
& & +\ \bigl(q_{1-\alpha}(X)+\dfrac{ g_n''(q_{1-\alpha}(X))}{
g_n'(q_{1-\alpha}(X))}\bigr) \Big(2\,u_j\,D_j\,g_n'(q_{1-\alpha}(X))- B_j\,
\sum_{i=1}^nu_i^2 D_i\Big)\bigg),\notag
\end{eqnarray}
with
\begin{align}
A_i&=\sqrt{\dfrac{\rho_i}{1-\rho_i}}\
\phi\left(\dfrac{c_i-\sqrt{\rho_i}\,q_{1-\alpha}(X)}{\sqrt{1-\rho_i}}\right)
\left(1-2\,\Phi\left(\dfrac{c_i-\sqrt{\rho_i}\,q_{1-\alpha}(X)}{\sqrt{1-\rho_i}}\right)\right),\nonumber\\
B_i&=-\phi\left(\dfrac{c_i-\sqrt{\rho_i}\,q_{1-\alpha}(X)}{\sqrt{1-\rho_i}}\right)\,
\sqrt{\dfrac{\rho_i}{1-\rho_i}},\nonumber\\
D_i&=\Phi\left(\dfrac{c_i-\sqrt{\rho_i}\,q_{1-\alpha}(X)}{\sqrt{1-\rho_i}}\right)-
\Phi\left(\dfrac{c_i-\sqrt{\rho_i}\,q_{1-\alpha}(X)}{\sqrt{1-\rho_i}}\right)^2,\nonumber\\
E_i&=\phi\left(\dfrac{c_i-\sqrt{\rho_i}\,q_{1-\alpha}(X)}{\sqrt{1-\rho_i}}\right)\,\dfrac{\rho_i}{1-\rho_i}
\,\dfrac{\sqrt{\rho_i}\,q_{1-\alpha}(X)-c_i}{\sqrt{1-\rho_i}}\,.\nonumber
\end{align}
\end{corollary}
Proposition \ref{pr:0} and Corollary \ref{co:1} provide a way for the
simultaneous determination of economic capital charges in a top-down approach.
First, by means of Proposition \ref{pr:0} an approximation for the portfolio
value-at-risk is calculated. Then, by means of Corollary \ref{co:1}
approximations for the partial derivatives of the portfolio value-at-risk can
be found. The derivatives are related to the capital charges via
\begin{equation}\label{eq:charge}
  \text{Capital charge of asset }j\ = \ u_j\,\frac{\partial q_\alpha(L_n)}{\partial
u_j}.
\end{equation}
However, there is a caveat in using this approach. Its accuracy is subject to
the validity of (\ref{eq:cond}) which will be questionable if there are very
large exposures in the portfolio. In this case, the semi-asymptotic approach
we will explain in Section \ref{sec:part} might be more appropriate.

\section{The semi-asymptotic approach}
\label{sec:part}
We consider here a special case of (\ref{eq:8}) where $\rho_1 = \tau$, $c_1 =
a$ but $\rho_i = \rho$ and $c_i = c$ for $i > 1$, and $\sum_{i=1}^n u_i = 1$.
Additionally, we assume that $u_1 = u$ is a constant for all $n$ but that
$u_2, u_3, \ldots$ fulfills (\ref{eq:cond}).

In this case, the portfolio loss can be represented by
\begin{equation}
  \label{eq:3.0}
  L_n(u, u_2, \ldots, u_n) \,=\, u\,\mathbf{1}_{\{\sqrt{\tau}\,X +
  \sqrt{1-\tau}\,\xi \le a\}} + (1-u) \sum_{i=2}^n u_i\,
  \mathrm{1}_{\{\sqrt{\rho}\,X + \sqrt{1-\rho}\,\xi_i\le c\}}
\end{equation}
with $\sum_{i=2}^n u_i = 1$. Transition to the limit for $n\to \infty$ in
(\ref{eq:3.0}) in the sense of \eqref{eq:cond} 
leads to the \emph{semi-asymptotic} percentage loss function
\begin{equation}
  \label{eq:3.1}
  L(u)\,=\,u\,\mathbf{1}_D + (1-u)\,Y
\end{equation}
with $D = \{\sqrt{\tau}\,X +
  \sqrt{1-\tau}\,\xi \le a\}$ and $Y = \mathrm{P}\bigl[\xi \le \frac{c-
  \sqrt{\rho}\,x}{\sqrt{1-\rho}}\bigr]\Big|_{x=X}$. Note that the transition
  from \eqref{eq:3.0} to \eqref{eq:3.1} is \emph{semi}-asymptotic only as
we keep the first term on the right-hand side of \eqref{eq:3.0} unchanged such
  that it reappears in \eqref{eq:3.1}.
Of course, a natural
  choice for $\tau$ might be $\tau = \rho$, the mean portfolio asset correlation.
\begin{definition}
  \label{de:1}
The quantity
\begin{equation*}
u\,\mathrm{P}[D\,|\,L(u) = q_\alpha(L(u))]
\end{equation*}
is called\/ \emph{semi-asymptotic capital charge} (at level $\alpha$) of the
loan with exposure $u$ (as percentage of total portfolio exposure) and default
event $D$ as in (\ref{eq:3.0}).
\end{definition}
The capital charges we suggest in  Definition \ref{de:1} have to be calculated
separately, i.e.\ for each asset an own model of type (\ref{eq:3.1}) has to be
regarded. This corresponds to a bottom-up approach since the total capital
requirement for the portfolio must be determined by adding up all the capital
charges of the assets. Note that the capital charges of Definition \ref{de:1}
are not portfolio invariant in the sense of \citet{Gordy01}. However, in
contrast to the portfolio invariant charges, the semi-asymptotic charges take
into account concentration effects. In particular, their dependence on the
exposure $u$ is not merely linear since also the factor $\mathrm{P}[D\,|\,L(u)
= q_\alpha(L(u))]$ depends upon $u$. Definition \ref{de:1} is in line with the
general definition of VaR contributions \citep[cf.][]{L96, Tasche99} since
(\ref{eq:3.1}) can be considered a two-assets portfolio model.

In the following, we will assume that the conditional distribution functions
$F_0$ and $F_1$ of $Y$ given $\mathbf{1}_D = 0$ and  $\mathbf{1}_D = 1$
respectively have densities which are continuous and concentrated on the
interval $(0,1)$. Call these densities $f_0$ and $f_1$, i.e.
\begin{equation}
  \label{eq:3.2}
  \mathrm{P}[Y \le y\,|\,\mathbf{1}_D = i]\,=\,F_i(y) \,=\, \int_{-\infty}^y
  f_i(t)\,d t, \quad y\in\mathbb{R},\ i = 0,1.
\end{equation}
In this case, the distribution function and the density of $L(u)$ are given by
\begin{subequations}
\begin{align}\label{eq:3.2a}
  \mathrm{P}[L(u) \le z] & = p\,F_1\bigl(\frac{z-u}{1-u}\bigl) +
  (1-p)\,F_0\bigl(\frac{z}{1-u}\bigl)\\
\intertext{and}
\frac{\partial \mathrm{P}[L(u) \le z]}{\partial z} & = (1-u)^{-1}\left( p\,f_1\bigl(\frac{z-u}{1-u}\bigl) +
  (1-p)\,f_0\bigl(\frac{z}{1-u}\bigl)\right)\label{eq:3.2b}
\end{align}
\end{subequations}
respectively, where $p = \mathrm{P}[D]$ is the \emph{default probability} of
the loan under consideration. By means of (\ref{eq:3.2a}) and  (\ref{eq:3.2b}), the quantile
  $q_\alpha(L(u))$ can be numerically computed. For the conditional
  probability which is part of Definition \ref{de:1}, we obtain
  \begin{equation}
    \label{eq:3.3}
 \mathrm{P}[D\,|\,L(u) = z]\,=\,
\frac{p\,f_1\bigl(\frac{z-u}{1-u}\bigl)}{ p\,f_1\bigl(\frac{z-u}{1-u}\bigl) +
  (1-p)\,f_0\bigl(\frac{z}{1-u}\bigl)}.
  \end{equation}
If, similarly to the situation of (\ref{eq:10}), we assume that $X$ and $\xi$
are independent and both standard normally distributed, we obtain for the
conditional densities in (\ref{eq:3.2})
\begin{subequations}
\begin{align}\label{eq:3.4a}
f_1(z) & =  \left\{
\begin{array}{l@{\quad}l}
\frac{\displaystyle f(z)}{\displaystyle p} \,\Phi\left( \left({\textstyle
\sqrt{1-\tau}}\right)^{-1} \bigl( \Phi^{-1}(p) + \sqrt{\tau}\,\bigl( \frac{%
\sqrt{1-\rho}\,\Phi^{-1}(z) - c}{\sqrt{\rho}}\bigr) \bigr) \right), & z
\in (0,1) \\
0, & \mbox{otherwise}
\end{array}
\right.\\
\intertext{and}
f_0(z) & =  \left\{
\begin{array}{l@{\quad}l}
\frac{\displaystyle f(z)}{\displaystyle 1- p} \,\Phi\left( - \left({\textstyle
\sqrt{1-\tau}}\right)^{-1} \bigl( \Phi^{-1}(p) + \sqrt{\tau}\,\bigl( \frac{%
\sqrt{1-\rho}\,\Phi^{-1}(z) - c}{\sqrt{\rho}}\bigr) \bigr) \right), & z
\in (0,1) \\
0, & \mbox{otherwise,}
\end{array}
\right.\label{eq:3.4b}\\
\intertext{with}
f(z) & =  \left\{
\begin{array}{l@{\quad}l}
\sqrt{\rho/(1-\rho)}\, \exp\left( 1/2 \left( \Phi^{-1}(z)^2 - \left( \frac{%
\sqrt{1-\rho}\,\Phi^{-1}(z) - c}{\sqrt{\rho}}\right)^2\right)\right), & z
\in (0,1) \\
0, & \mbox{otherwise.}
\end{array}
\right.
\end{align}
\end{subequations}
Note that $p = \Phi(a)$ in (\ref{eq:3.4a}) and (\ref{eq:3.4b}). Denote by
$\Phi_2(\cdot, \cdot;\theta)$ the distribution function of the bivariate standard
normal distribution with correlation $\theta$. Then, for the
conditional distribution functions corresponding to (\ref{eq:3.4a}) and
(\ref{eq:3.4b}),
we have
\begin{subequations}
\begin{align}\label{eq:F1}
  F_1(z) &=
  \begin{cases}
     1 - p^{-1}\, \Phi_2\bigl(a, \frac{c -
    \sqrt{1-\rho}\,\Phi^{-1}(z)}{\sqrt{\rho}}; \sqrt{\tau}\bigr), & z
\in (0,1) \\
0, & \mbox{otherwise,}
  \end{cases}\\
\intertext{and}
  F_0(z) &=
  \begin{cases}
    (1- p)^{-1}\, \Phi_2\bigl(- a,- \frac{c -
    \sqrt{1-\rho}\,\Phi^{-1}(z)}{\sqrt{\rho}}; \sqrt{\tau}\bigr), & z
\in (0,1) \\
0, & \mbox{otherwise.}
  \end{cases}\label{eq:F0}
\end{align}
\end{subequations}
Unfortunately, the capital charges from Definition \ref{de:1} do not yield reasonable values
in all cases. Indeed, considering the relation of the default probability $p$
and the confidence level $\alpha$, we easily arrive at the following result.
\begin{proposition}
  \label{pr:1}
For $L(u)$ defined by (\ref{eq:3.1}) with $F_1(0) = 0$, we have
\begin{equation*}
  \lim_{u\to 1} q_\alpha(L(u)) \,=\,
  \begin{cases}
    1, & p > 1- \alpha\\
    0, & p \le 1 - \alpha.
  \end{cases}
\end{equation*}
\end{proposition}
The proof of Proposition \ref{pr:1} is presented in Appendix
\ref{sec:appendix}.
Under slightly stronger assumptions than in Proposition \ref{pr:1}, the
limiting behavior of  $q_\alpha(L(u))$ is inherited by
$\mathrm{P}[D\,|\,L(u)=q_\alpha(L(u))]$.
\begin{proposition}
  \label{pr:2}
Assume that the densities in (\ref{eq:3.2}) are positive in $(0,1)$. Then for
$L(u)$ defined by (\ref{eq:3.1}), we have
\begin{equation*}
  \lim_{u\to 1}  \mathrm{P}[D\,|\,L(u)=q_\alpha(L(u))]\,=\,
  \begin{cases}
    1, & p > 1- \alpha\\
    0, & p < 1 - \alpha.
  \end{cases}
\end{equation*}
\end{proposition}
The proof of Proposition \ref{pr:2} is also given in Appendix
\ref{sec:appendix}. In addition, Remark \ref{rm:1} explains why there is no
statement on the case $\alpha = 1-p$ in Proposition \ref{pr:2}.

By Proposition \ref{pr:2}, we see that in case $p < 1 -
\alpha$ it would be worthwhile to concentrate all the exposure to one loan.
This is another appearance of a well-known deficiency of value-at-risk. Under
value-at-risk portfolio risk measurement, putting all the risk into an event
with very small probability can quite drastically reduce capital charges. In
order to avoid this phenomenon, other risk measures like Expected Shortfall
\citep[cf.][]{AT02} have to be used. For Expected Shortfall, in Definition
\ref{de:1} the conditional probability $\mathrm{P}[D\,|\,L(u) =
q_\alpha(L(u))]$ has to be replaced by  $\mathrm{P}[D\,|\,L(u) \ge
q_\alpha(L(u))]$. See \citet{TascheTheiler} for details of this approach.

\section{Numerical example}
\label{sec:num}
\setcounter{figure}{0}
For the purpose of illustrating the previous sections, we consider a numerical
example in the framework of Section \ref{sec:part}. This means that there is a
portfolio driven by systematic risk only (the variable $Y$ in (\ref{eq:3.1}))
which is enlarged with an additional loan (the indicator $\mathbf{1}_D$ in
(\ref{eq:3.1})).


The portfolio modeled by $Y$ has a quite moderate credit standing which is
expressed by its expected loss $\mathrm{E}[Y] = 0.025 = \Phi(c)$. However, we
assume that the additional loan enjoys a quite high credit-worthiness as we
set $p = \mathrm{P}[D] = 0.002 = \Phi(a)$. The asset correlation $\rho$ of the
portfolio with the systematic factor and the asset correlation $\tau$ of the
additional loan with the systematic factor are chosen according to the current
Basel~II proposal for corporate loan portfolios \citep{BC04}, namely $\rho =
0.154$ and $\tau = 0.229$.

In order to draw Figure \ref{fig:1}, the risk of the portfolio loss variable
$L(u)$, as defined by (\ref{eq:3.1}), is regarded as a function of the
relative weight $u$ of the new loan in the portfolio. Risk then is calculated
as true Value-at-Risk (VaR) at level 99.9\% (i.e.\ $q_{0.999}(L(u))$), as the
granularity adjustment approximation to $q_{0.999}(L(u))$ according to
(\ref{eq:5}), and in the manner\footnote{%
Actually, the regulatory capital requirements according to Basel II are to be calculated as
\emph{unexpected losses}, i.e.\ as differences of quantiles and loss
expectations. In this example, risk is calculated as the quantile only, i.e.\
as sum of expected and unexpected loss in the Basel II sense.
} of the Basel~II Accord \citep[cf.][]{BC04}.
The values of VaR are computed by solving numerically for $z$ the equation
\begin{equation}\label{eq:num}
  \mathrm{P}[L(u) \le z]\,=\, \alpha,
\end{equation}
with $\mathrm{P}[L(u) \le z]$ given by (\ref{eq:3.2a}), (\ref{eq:F1}), and
(\ref{eq:F0}). The granularity adjustment approximation is computed by
applying (\ref{eq:14}) to (\ref{eq:3.0}) and taking the limit with
$n\to\infty$. The Basel~II risk weight function can be interpreted as the
first order version of (\ref{eq:5}) which yields the straight line
\begin{equation}\label{eq:Basel_II}
  u \, \mapsto\,
  u\,\Phi\left(\frac{a-\sqrt{\tau}\,q_{1-\alpha}(X)}{\sqrt{1-\tau}}\right) +
  (1-u)\,\Phi\left(\frac{c-\sqrt{\rho}\,q_{1-\alpha}(X)}{\sqrt{1-\rho}}\right).
\end{equation}
Note that for (\ref{eq:Basel_II}) we make use of a stylized version of the
Basel~II risk weight function as we chose arbitrarily the correlation $\rho$.
The Basel~I risk weight function is a constant at 8\% that should not be
compared to the afore-mentioned functions since it was defined without
recourse to mathematical methods. For that reason it has been omitted from
Figure~\ref{fig:1}.
\refstepcounter{figure}
\ifpdf
\begin{figure}[ht]
  \centering
  \resizebox{\height}{10.0cm}{\includegraphics[width=10.0cm]{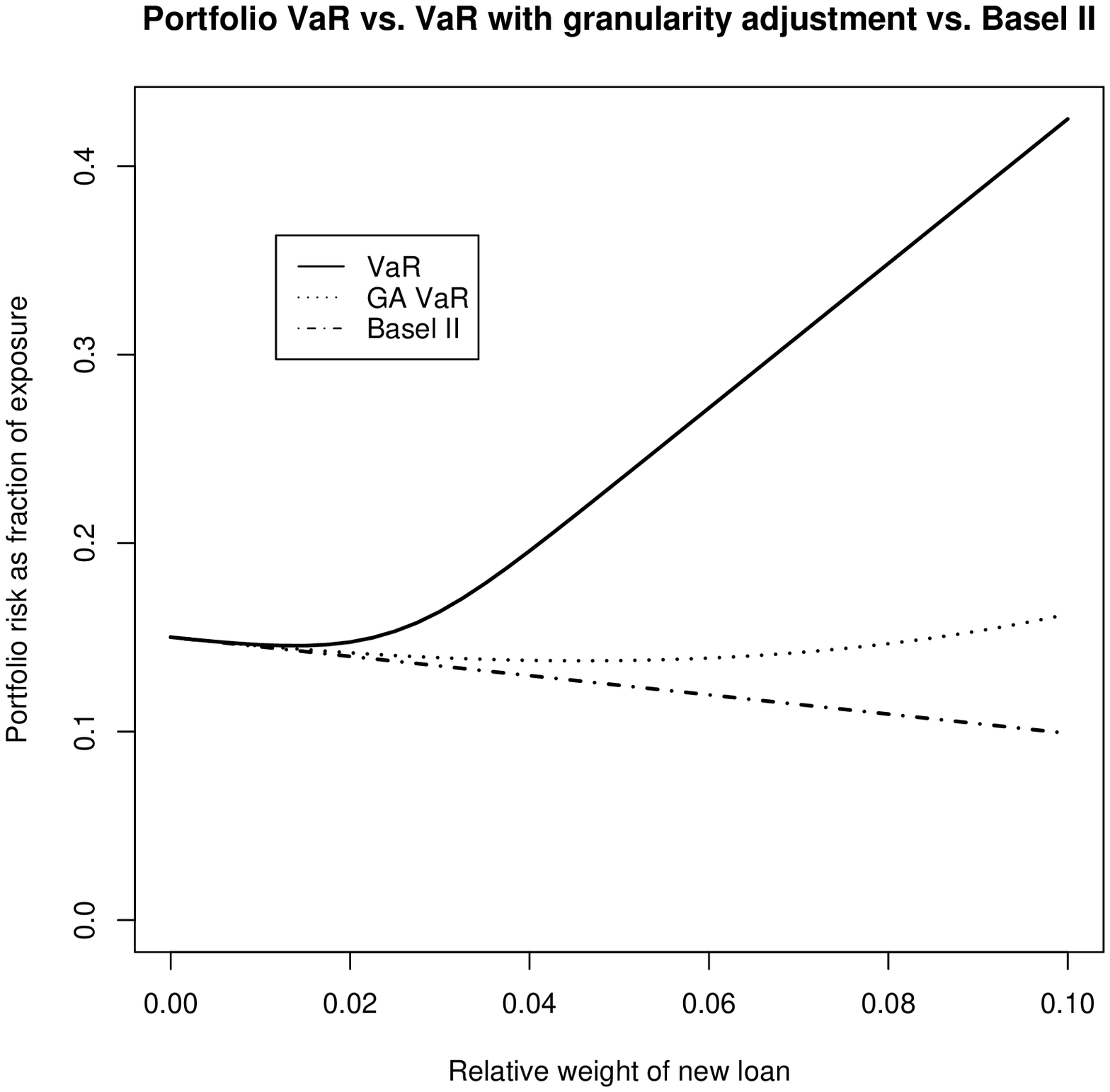}}
  \parbox{12.0cm}{\footnotesize{}Figure \thefigure:
Portfolio risk as function of the relative weight of the additional loan
  in the setting of Section \ref{sec:part}. Comparison of true Value-at-Risk (VaR),
  granularity adjustment approximation to VaR (GA VaR), and the risk weights according to
  the Basel~II Accord.}\label{fig:1}
\end{figure}
\else
\begin{figure}[ht]
  \centering
\resizebox{\height}{10.0cm}{
\includegraphics[width=10.0cm]{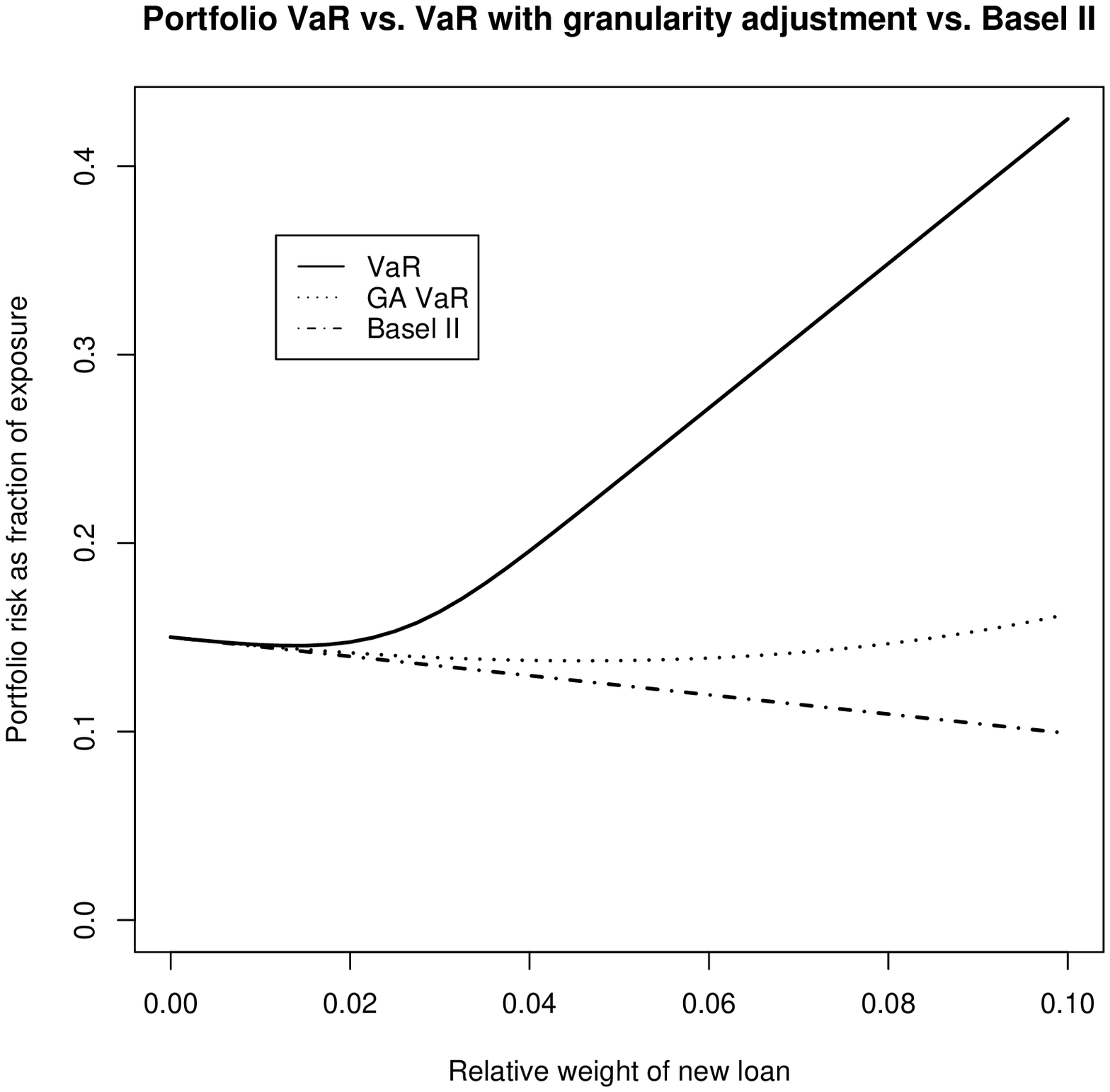}}
  \parbox{12.0cm}{\footnotesize{}Figure \thefigure:
Portfolio risk as function of the relative weight of the additional loan
  in the setting of Section \ref{sec:part}. Comparison of true Value-at-Risk (VaR),
  granularity adjustment approximation to VaR (GA VaR), and the risk weights according to
  the Basel~II Accord.}\label{fig:1}
\end{figure}
\fi
From Figure~\ref{fig:1}, we see that, up to a level of 2\% relative weight of
the new loan, both the granularity adjustment approach and the Basel~II
approach yield precise approximations to the true portfolio VaR. Beyond that
level the approximations lose quality very fast. Of course, this observation
does not surprise since both the granularity adjustment approach and the
Basel~II are based on the assumption that there is no concentration in the
portfolio. Intuitively, this condition should be violated when the
concentration of the additional loan increases too much.
\refstepcounter{figure}
\ifpdf
\begin{figure}[ht]
  \centering
  \resizebox{\height}{10.0cm}{\includegraphics[width=10.0cm]{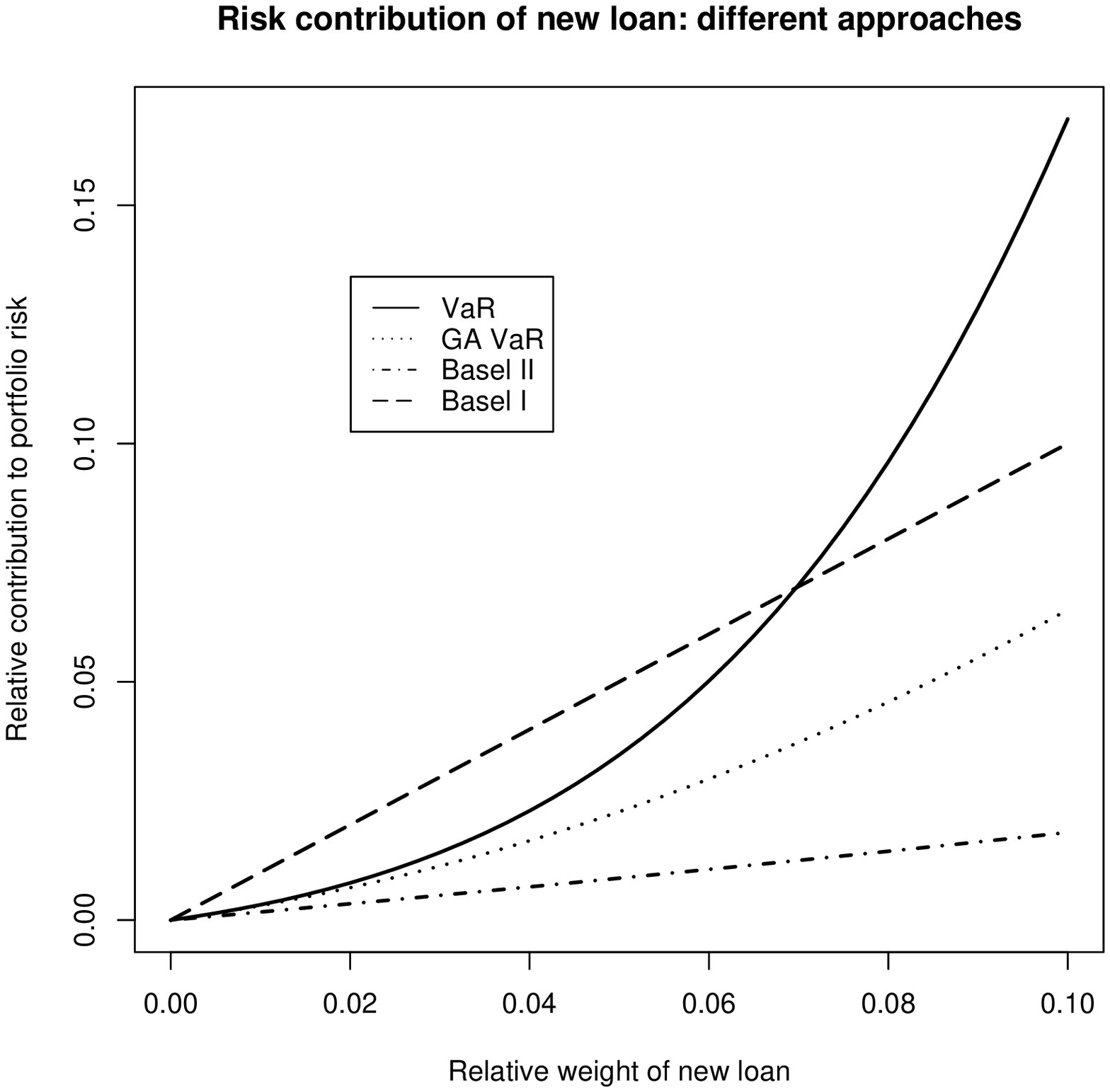}}
  \parbox{12.0cm}{\footnotesize{}Figure \thefigure:
Relative risk contribution of new loan as function of the relative weight of the new loan
  in the setting of Section \ref{sec:part}. Comparison of contribution to true Value-at-Risk (VaR),
  contribution to granularity adjustment approximation to VaR (GA VaR),
and the contributions according to
  the Basel~II and Basel~I Accords.}\label{fig:2}
\end{figure}
\else
\begin{figure}[ht]
  \centering
\resizebox{\height}{10.0cm}{
\includegraphics[width=10.0cm]{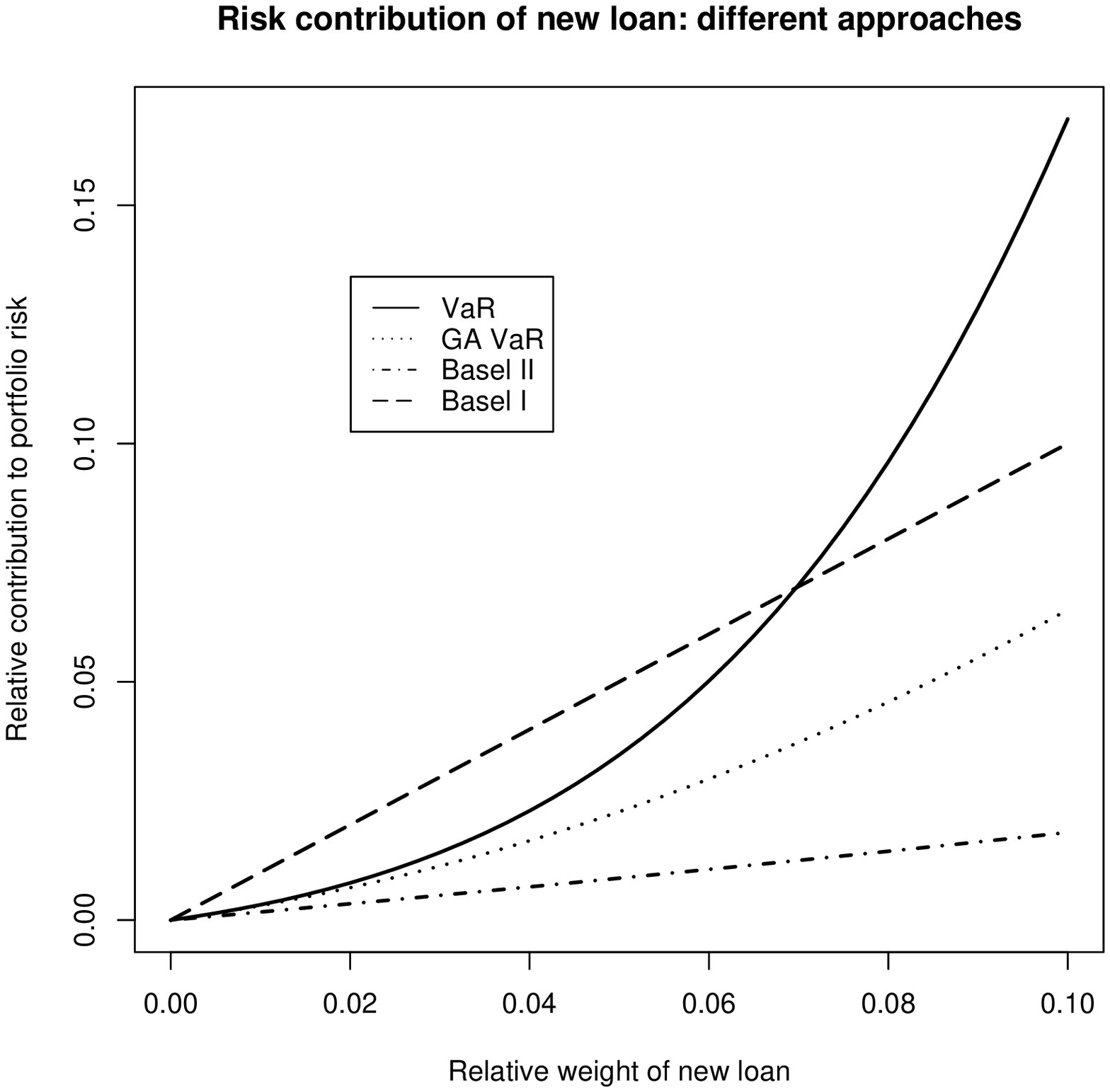}}
  \parbox{12.0cm}{\footnotesize{}Figure \thefigure:
Relative risk contribution of new loan as function of the relative weight of the new loan
  in the setting of Section \ref{sec:part}. Comparison of contribution to true Value-at-Risk (VaR),
  contribution to granularity adjustment approximation to VaR (GA VaR), and the contributions according to
  the Basel~II and Basel~I Accords.}\label{fig:2}
\end{figure}
\fi

Figure~\ref{fig:2} illustrates the relative contribution of the new loan to
the risk of the portfolio loss variable $L(u)$. The contribution is considered
a function of the relative weight $u$ of the new loan in the portfolio and
calculated according to four different methods. The first method relates to
the relative contribution to true portfolio VaR, defined as the ratio of the
contribution to VaR according to Definition~\ref{de:1} and portfolio VaR, i.e.\
the function
\begin{equation}\label{eq:me1}
  u\,\mapsto\,\frac{u\,\mathrm{P}[D\,|\,L(u) =
  q_\alpha(L(u))]}{q_\alpha(L(u))},
\end{equation}
where the conditional probability has to be evaluated by means of
(\ref{eq:3.3}), (\ref{eq:3.4a}), and (\ref{eq:3.4b}). Secondly, the relative
contribution is calculated as the fraction with the granularity adjustment
contribution according to Corollary \ref{co:1} applied to (\ref{eq:3.0}) (with
$n\to\infty$) in
the numerator and the granularity adjustment approximation to VaR as for
Figure~\ref{fig:1} in the denominator. Moreover, curves are drawn for the
Basel~II approach, i.e.\ the function
\begin{equation}\label{eq:contribs}
  u \, \mapsto\,
  \frac{u\,\Phi\left(\frac{a-\sqrt{\tau}\,q_{1-\alpha}(X)}{\sqrt{1-\tau}}\right)}
  {u\,\Phi\left(\frac{a-\sqrt{\tau}\,q_{1-\alpha}(X)}{\sqrt{1-\tau}}\right) +
  (1-u)\,\Phi\left(\frac{c-\sqrt{\rho}\,q_{1-\alpha}(X)}{\sqrt{1-\rho}}\right)},
\end{equation}
and the Basel~I approach. The latter approach just entails the diagonal as
risk contribution curve since it corresponds to purely volume-oriented capital
allocation.

Note that in Figure~\ref{fig:2} the true VaR curve intersects the diagonal
(Basel~I curve) just at the relative weight $u^\ast$ that corresponds to the
minimum risk portfolio $L(u^\ast)$. Whereas the granularity adjustment curve
yields, up to a relative risk weight of 2\% of the additional loan, a
satisfying approximation to the true contribution curve, the Basel~II curve
loses reliability much earlier. Both these approximation curves are completely
situated below the diagonal. This fact could yield the misleading impression
that an arbitrarily high exposure to the additional loan still improves the
risk of portfolio. However, as the true VaR curve in Figure~\ref{fig:2} shows,
the diversification effect from pooling with the new loan stops at 7\%
relative weight. From Figure~\ref{fig:1} we see that beyond 7\% relative
weight of the additional loan the portfolio risk grows dramatically.

\section{Conclusions}
\label{sec:concl}
In the present paper we have extended the granularity approach introduced by
\citet{Gordy01} in a way that takes care of the exact weights of the assets in
the portfolio. By calculating the partial derivatives of the resulting
approximate portfolio loss quantile and multiplying the results with the asset
weights, we can arrive at an approximate capital charge for every asset in the
portfolio. In addition, we have introduced the semi-asymptotic approach to
one-factor portfolio loss modeling. This approach provides an alternative way
to capital charges that accounts for concentrations. By a numerical comparison
of these both approaches we have shown that the granularity approach to risk
contributions may come up with misleading results if there is a non-negligible
concentration in the portfolio. However, also the semi-asymptotic approach
yields counter-intuitive results if the probability of default of the loan
under consideration is very small. In order to avoid this drawback which is
caused by the very nature of value-at-risk, sounder risk measures like
Expected Shortfall have to be used.

\setcounter{section}{0} \renewcommand{\thesection}{\Alph{section}}
\section{Appendix}
\label{sec:appendix}

The proofs of Propositions \ref{pr:1} and \ref{pr:2} are based on the
following simple inequalities for $q_\alpha\bigl(L(u)\bigr)$.
\begin{lemma}
  \label{le:1}
Assume $F_1(0) = 0$. Then
for $L(u)$ defined by (\ref{eq:3.1}), we have
\begin{equation*}
q_\alpha\bigl(L(u)\bigr)
\begin{cases}
  \le 1-u, & \mbox{if } \alpha \le 1-p,\\
> u, & \mbox{if } \alpha > 1-p.
\end{cases}
\end{equation*}
\end{lemma}
\textbf{Proof.}
In case $\alpha \le 1-p$ we obtain by means of (\ref{eq:3.2a})
\begin{equation}\label{eq:A.1}
  \mathrm{P}[L(u) \le 1-u] \,=\, (1-p)\,F_0\bigl(\frac{1-u}{1-u}\bigr) + p\,
F_1\bigl(\frac{1-2\,u}{1-u}\bigr)\,\ge\, 1-p\,\ge\,\alpha.
\end{equation}
By the definition of quantiles (see (\ref{eq:4})), (\ref{eq:A.1}) implies the
first part of the assertion. Assume now $\alpha > 1-p$. Analogously to
(\ref{eq:A.1}), then we calculate
\begin{equation}
  \label{eq:A.2}
\mathrm{P}[L(u) \le u] \,=\, (1-p)\,F_0\bigl(\frac{u}{1-u}\bigr) + p\,
F_1(0) \,\le \,1-p\, < \,\alpha.
\end{equation}
Again by the definition of quantiles,  (\ref{eq:A.2}) yields $q_\alpha(L(u)) >
u$ as stated above.
\hfill $\Box$

\textbf{Proof of Proposition \ref{pr:1}.} In case $\alpha \le 1-p$,
Lemma \ref{le:1} implies
\begin{subequations}
  \begin{equation}
    \label{eq:A.4}
    \limsup_{u\to 1} q_\alpha\bigl(L(u)\bigr) \,\le \,0,
  \end{equation}
in case $\alpha > 1-p$ we deduce from Lemma \ref{le:1}
\begin{equation}
  \label{eq:A.5}
  \liminf_{u\to 1} q_\alpha\bigl(L(u)\bigr) \,\ge \,1.
\end{equation}
\end{subequations}
Since $0\le q_\alpha\bigl(L(u)\bigr) \le 1$, (\ref{eq:A.4}) and (\ref{eq:A.5})
imply the assertion.\hfill $\Box$

\textbf{Proof of Proposition \ref{pr:2}.}
By Lemma  \ref{le:1}, in case  $\alpha \le 1-p$ we have
$q_\alpha\bigl(L(u)\bigr) \le 1/2$ for $u > 1/2$ and hence
$q_\alpha\bigl(L(u)\bigr) -u < 0$. This implies
\begin{equation}
  \label{eq:A.6}
  f_1\bigl(\frac{q_\alpha(L(u)) -u}{1-u}\bigr) \,=\,0
\end{equation}
for $u > 1/2$. A closer inspection of the proof of Lemma \ref{le:1} reveals
that in case $\alpha < 1-p$  even $0 < q_\alpha(L(u)) < 1-u$  obtains if $f_0$
is positive. This observation implies
$f_0\bigl(\frac{q_\alpha(L(u))}{1-u}\bigr) >0$. Hence, $\lim_{u\to 1}
\mathrm{P}[D\,|\,L(u)=q_\alpha(L(u))] = 0$ follows from (\ref{eq:3.3}) and
(\ref{eq:A.6}). Assume now $\alpha > 1-p$. Then by the positivity assumption
on $f_1$, Lemma \ref{le:1}  implies $1 > q_\alpha\bigl(L(u)\bigr) > u$. Since
$u / (1-u) > 1$ for $u > 1/2$ we may deduce that
\begin{equation}
  \label{eq:A.8}
f_0\bigl(\frac{q_\alpha(L(u))}{1-u}\bigr)\, =\, 0  \quad \text{and} \quad
f_1\bigl(\frac{q_\alpha(L(u))-u}{1-u}\bigr) \, > \,0.
\end{equation}
This implies $\mathrm{P}[D\,|\,L(u)=q_\alpha(L(u))] = 1$ for $u$ enough close
to $1$. \hfill $\Box$
\begin{remark}\label{rm:1}
By inspection of the proof of Proposition \ref{pr:2}, one can observe that
\begin{equation}
  \label{eq:rem}
q_{1-p}(L(u))  \,=\,1-u, \qquad u > 1/2.
\end{equation}
As a consequence, $f_1\bigl(\frac{q_{1-p}(L(u))-u}{1-u}\bigr)$ and
$f_0\bigl(\frac{q_{1-p}(L(u))}{1-u}\bigr)$ may equal zero at the same time.
Hence, in case $\alpha = 1-p$ according to (\ref{eq:3.3}) the conditional
probability $\mathrm{P}[D\,|\,L(u)=q_{1-p}(L(u))]$ may be undefined for $u >
1/2$. For this reason, Proposition \ref{pr:2} does not provide a limit
statement for the case  $\alpha = 1-p$.
\end{remark}


\end{document}